\documentclass[%
reprint,
preprintnumbers,
 amsmath,amssymb,
 aps,
]{revtex4-2}
\usepackage{tabularx}
\usepackage{multirow}
\usepackage{subfig}
\newcommand{\I}{$\mathrm{i}$}
\usepackage{graphicx}
\usepackage{dcolumn}
\usepackage{bm}
\usepackage[colorlinks, citecolor=blue]{hyperref}

\usepackage{tikz}
\usetikzlibrary{decorations.pathmorphing}
\usetikzlibrary{decorations.markings}
\usetikzlibrary{positioning, shapes, snakes, arrows}

\tikzset{
	graviton/.style={line width=.8pt, decorate, decoration={snake, segment length=4pt,amplitude=1.8pt, pre length=.1cm, post length=.0cm}},
	worldline/.style={gray, line width=1pt},
	worldlineBold/.style={black, line width=.6pt},
	zUndirected/.style={line width=1pt},
	zParticle/.style={line width=1pt,postaction={decorate},decoration={markings,mark=at position .6 with {\arrow[#1]{latex}}}},
	zParticleF/.style={line width=1pt,postaction={decorate}},
	cscalar/.style={line width=1pt,postaction={decorate},decoration={markings,mark=at position .6 with {\arrow[#1]{latex}}}},
	cscalar2/.style={line width=1pt,postaction={decorate},decoration={markings,mark=at position .8 with {\arrow[#1]{latex}}}},
	photon/.style={line width =.8pt, decorate, decoration={snake, segment length=4pt, amplitude=1.8pt,  pre length=.1cm, post length=.1cm}}
}
\newcommand{\ads}[1]{$\mathrm{AdS}_{#1}$}
\newcommand{\RN}{Reissener-Nordstr\"{o}m}
\newcommand{\JT}{Jackiw-Teitelboim}
\usepackage[justification=raggedright,singlelinecheck=true]{caption}
\begin{document}

\preprint{PCFT-25-25}

\title{Quantum Gravity Corrections to the Scalar Quasi-Normal Modes in Near-Extremal Reissener-Nordstr\"{o}m Black Holes}

\author{Zheng Jiang$^{1,2}$}
\email{jiangzheng22@mails.ucas.ac.cn}
\author{Jun Nian$^{1,3}$}
\email{nianjun@ucas.ac.cn}
\author{Caiying Shao$^2$}
\email{shaocaiying@ucas.ac.cn}
\author{Yu Tian$^{1,2}$}
\email{ytian@ucas.ac.cn}
\author{Hongbao Zhang$^{4,5}$}
\email{hongbaozhang@bnu.edu.cn}
\affiliation{$^1$International Centre for Theoretical Physics Asia-Pacific,\\
University of Chinese Academy of Sciences, 100190 Beijing, China\\
$^2$School of Physical Sciences, University of Chinese Academy of Sciences, 100049 Beijing, China\\
$^3$Peng Huanwu Center for Fundamental Theory, Hefei, Anhui 230026, China\\
$^4$School of Physics and Astronomy, Beijing Normal University, Beijing 100875, China\\
$^5$Key Laboratory of Multiscale Spin Physics, Ministry of Education, Beijing Normal University, Beijing 100875, China}


\begin{abstract}
We investigate quantum corrections to scalar quasi-normal modes (QNMs) in the near-extremal Reissner–Nordstr\"om black hole background with quantum correction in the near-horizon AdS$_2\times \mathrm{S}^2$ region. By performing a dimensional reduction, we obtain an effective Jackiw–Teitelboim (JT) gravity theory, whose quantum fluctuations are captured by the Schwarzian action. Using path integral techniques, we derive the quantum-corrected scalar field equation, which modifies the effective potential governing the QNMs. These corrections are extended from the near-horizon region to the full spacetime via a matching procedure. We compute the corrected QNMs using both the third-order WKB method and the Prony method and find consistent results. Our analysis reveals that quantum corrections can lead to substantial shifts in the real parts of QNM frequencies, particularly for small-mass or near-extremal black holes, while the imaginary parts remain relatively stable. This suggests that quantum gravity effects may leave observable imprints on black hole perturbation spectra, which could be potentially relevant for primordial or microscopic black holes.
\end{abstract}

\maketitle


\section{\label{sec:Introduction}Introduction}
The detection of gravitational waves has opened a new observational window to the universe \cite{GWII}. Among the various stages of a gravitational wave signal, the ringdown phase, which is dominated by the quasi-normal modes (QNMs) of the remnant black hole, provides a particularly clean and well-understood probe of the underlying spacetime geometry.

Black holes should not be viewed as isolated systems in real physics problems \cite{Konoplya_2011}. Especially interesting in this regard is the linear perturbation theory. If the perturbation is performed, the late-time behavior should be a superposition of many exponentially-damped modes, whose complex frequencies $\omega$ are properties of the black hole, independent of the initial perturbation. These complex frequencies are then referred to as QNMs, which have attracted considerable interest. Considering the quantized gravitational field, the QNMs of the perturbative fields should be corrected. Quantum corrections to the QNMs by the UV completion of the gravity have been discussed recently in \cite{koshelev2024}, which added higher-curvature terms suppressed by the Planck mass $M_{\text{pl}}$ to the action. In contrast, in this paper, we will focus on the Schwarzian action, which is suppressed by a new energy scale $C^{-1}$ related to the boundary dilaton (see Eq.~\eqref{eq:partition in C}). We will see that $C^{-1}$ is much smaller than the Planck mass $M_{\text{pl}}$ \eqref{eq: evaluation of C}. Hence, the results of \cite{koshelev2024} correspond to the UV completion of gravity, while our results correspond to the IR corrections, as also stated in \cite{maldacena2016}.

Thermal fluctuations are suppressed in the low temperature regime, due to the Boltzmann factor $\textrm{exp}(-1/ k_B T)$. However, quantum fluctuations, which originate from the Heisenberg uncertainty principle and persist even at zero temperature, become dominant in the dynamical behavior of the system. Hence, in the near-extremal limit, the temperature is low and the quantum fluctuations become important. In our paper, we investigate the quantum corrections to the scalar quasi-normal modes due to the quantum fluctuations of the near-extremal RN black holes.

For extremal and near-extremal RN black holes, the near-horizon region is \ads{2}$\times\mathrm{S}^2$. Upon dimensional reduction to the near-horizon AdS$_2$ throat region, we obtain an \ads{2} Jackiw-Teitelboim gravity \cite{Teitelboim:1983ux, Jackiw:1984je}. According to the \ads{}/CFT correspondence, the dual theory of the two-dimensional Jackiw-Teitelboim (JT) gravity is the one-dimensional Schwarzian theory.

Over the last decade, the \ads{2}-JT gravity/Schwarzian theory duality has garnered significant research interest, and considerable progress has been made. First, the backreaction of the dilaton gravity in \ads{2} was considered in \cite{almheiri2015}. For the \ads{2}-JT theory, the bulk action is a dilaton field coupled to the Einstein-Hilbert action. As discussed in \cite{maldacena2016}, the IR theory of gravity can be determined by the boundary cutoff. If we fix the proper length of the cutoff curve, we will see that the quantum fluctuations of the \ads{2} metric can be viewed as the reparametrization of the cutoff curve. Integrating out the bulk part of the \ads{2}-JT gravity, the boundary extrinsic curvature is shown to be proportional to the Schwarzian derivative with respect to the reparametrization modes, which are consequently named the Schwarzian modes.

We can expand the fluctuating \ads{2} metric and the physical quantities associated with it in terms of the Schwarzian modes. By introducing an auxiliary field, \cite{Stanford_2017} shows that the path integral of the Schwarzian theory is one-loop exact. The correlators of the Schwarzian modes at the one-loop level are obtained in \cite{Qi_2019}. Hence, using the Schwarzian action, we obtain the quantum gravity corrections of physical quantities if they are minimally coupled to the \ads{2}-JT gravity.

Some previous work along this line includes the quantum correction to geometric physical quantities \cite{Blommaert:2019hjr}, black hole evaporation in 2d JT gravity \cite{Mertens:2019bvy}, and defects in 2d JT gravity \cite{Mertens:2019tcm}, among others. While most of these papers treat the path integral of 2d JT gravity or 1d Schwarzian theory directly, in this paper, we evaluate the correlators perturbatively.

More specifically, to discuss the physical quantities in the near-extremal RN black hole background, we first need to perform a dimensional reduction. As discussed in \cite{Iliesiu_2021,Nayak_2018}, the near-horizon region of a near-extremal RN black hole can yield \ads{2}-JT gravity. The Euclidean path integral of such an effective theory can be computed exactly by first integrating out the gauge degrees of freedom \cite{iliesiu20192}, which also applies for the neutral scalar fields considered in this paper, and then by analyzing the Schwarzian modes. Hence, calculating quantum corrections in the near-horizon region involves finding the Schwinger-Dyson equation for the Schwarzian modes. For the far region, the quantum fluctuation should be small, which makes the far region essentially classical. However, there is an overlapping region between the near-horizon and far regions, which can carry quantum corrections from the near-horizon region to the far region. From this viewpoint, we can extend the Schwinger-Dyson equation from the near-horizon region to the whole RN spacetime. Similar ideas have been applied to compute the quantum-corrected greybody factor of the Hawking radiation for various near-extremal black holes \cite{brown2024eva, Maulik:2025hax, Lin:2025wof}. Moreover, \cite{li2025} considered the effect of counting the Schwarzian modes in decoherence and found a quantum gravity correction to the decoherence rate. In contrast to these works, we will study the quantum-corrected matter field equations in this paper.

The plan of this paper is as follows. In Sec.~\ref{sec:Quantum Corrections}, we will discuss how to introduce quantum corrections to the physical quantities in the near-extremal RN black hole background. In Sec.~\ref{sec:Scalar QNMs}, we will show how to calculate the scalar quasi-normal modes. In Sec.~\ref{sec:results}, we will list the results of quantum corrections to the scalar quasi-normal modes. Some discussions and outlooks are presented in Sec.~\ref{sec:discussion}.

\section{\label{sec:Quantum Corrections}Quantum correction to \ads{2} spacetime and RN black hole}

\subsection{\label{subsec:Rn to ads2}From \RN~Black Hole to \ads{2} Spacetime}
For the RN black hole, the metric is given by
\begin{equation}
    ds^2=f(r)dt^2+\frac{dr^2}{f(r)}+r^2d\theta^2+r^2\sin^2\theta d\varphi^2\, ,
\end{equation}
where the function $f(r)$ is given by the mass $M$ and the charge $Q$ of the black hole, as 
\begin{equation}
    f(r)=1-\frac{2M}{r}+\frac{Q^2}{r^2}\,.
\end{equation}

There are two horizons for the RN black hole,
\begin{equation}
    r_\pm=M\pm\sqrt{M^2-Q^2}\,,
\end{equation}
which are called the inner and outer horizons, respectively. To avoid the naked singularity, we require that the charge is no larger than the mass, i.e., $|Q|\leq M$. The Hawking temperature of an RN black hole is given as
\begin{equation}
    T=\frac{r_+-r_-}{8\pi Mr_+}=\frac{\sqrt{M^2-Q^2}}{4\pi M(M+\sqrt{M^2-Q^2})}\,.\label{eq:Hawking}
\end{equation}

When the black hole is extremal, $M=|Q|$, the two horizons coincide at $r_h$,
\begin{equation}
    r_+=r_-=r_h=M.
\end{equation}
While for the near-extremal black hole, two horizons come very close,
\begin{equation}
    r_\pm=r_h\pm \delta\,,\quad \delta=\sqrt{M^2-Q^2}\ll M\,.
\end{equation}
Considering the near-horizon region of an extremal RN black hole, i.e., $\frac{r-r_h}{r_h}\ll 1$, the metric can be written as
\begin{equation}
    ds^2=-\frac{R^2-\delta^2}{r_h^2}dt^2+\frac{r_h^2}{R^2-\delta^2}dR^2+r_h^2d\Omega_2^2\,,
\end{equation}
where $R=r-r_h$ and $d\Omega_2^2=d\theta^2+\sin^2\theta d\varphi^2$. Performing the coordinate transformation, $R=\delta\coth \delta z,T=t/r_h^2$, then the near-horizon region becomes \ads{2}$\times\mathrm{S}^2$ spacetime
\begin{equation}
    ds^2=r_h^2\delta^2\frac{dT^2+dz^2}{\sinh^2 \delta z}+r_h^2d\Omega_2^2\,.
\end{equation}
It is obvious that $\delta$ is proportional to the temperature, hence we can define $\delta={2\pi}/\beta$ to introduce the finite-temperature \ads{2}, as
\begin{equation}
    ds^2=r_h^2\frac{4\pi^2}{\beta^2}\frac{dT^2+dz^2}{\sinh^2(2\pi z/\beta)}\,.\label{eq:ads2fromrn}
\end{equation}

For the Einstein-Maxwell system, it has the action:
\begin{equation}
    I=\frac{1}{16\pi G}\int d^4x \sqrt{-g}(R-F_{\mu\nu}F^{\mu\nu})\,.
\end{equation}
Doing the dimensional reduction as in \cite{Iliesiu_2021}, and integrating out the Maxwell field, which does not interact with the scalar field considered in this paper, we will get the two-dimensional action:
\begin{widetext}
\begin{equation}
    I_{\mathrm{2d}}=\frac{\Phi_0}{16\pi G}\left(\int_{M_2} d^2x\sqrt{g^{2d}}R^{2d}-2\int_{\partial M_2}dt\sqrt{h} K \right)-\frac{1}{16\pi G}\left(\int_{M_2} d^2x\sqrt{g^{2d}}\Phi(R^{2d}+2)-2\int_{\partial M_2}dt\sqrt{h}\Phi_b K\right)\,,
\end{equation}
\end{widetext}
where the $\Phi_0+\Phi$ is the total dilaton. $\Phi_0$ is a constant and is of leading order, while $\Phi$ is of first order, and the $\Phi_b$ is the boundary dilaton. $R^{2d}$ is the scalar curvature and $K$ represents the extrinsic curvature. The first term is purely topological, according to the Gauss-Bonnet theorem. The second term is the \JT (JT) gravity, where $\Phi$ is called the dilaton.

From \ads{}/CFT correspondence\cite{Maldacena_1999,witten1998}, the \ads{2} spacetime has a one-dimensional dual theory. For the \ads{2}-JT gravity, the dual theory is the Schwarzian action, as discussed in the next subsection. For the \ads{2}-JT gravity, we can first do the path integral of the dilaton field, leading to the fixed \ads{} radius. For a fixed \ads{} radius spacetime, the metric can be solved as two functions, and the quantum fluctuation of the metric appears in these functions. However, we can also use another perspective to regard these functions as coordinate transformations. But according to the \ads{}/CFT correspondence, the quantum gravity of the \ads{2} spacetime at IR should be determined by a cutoff at the boundary. In the passive perspective (the latter one), we can regard the quantum fluctuation of the metric as a reparametrization of the boundary curve.

\subsection{\label{subsec:Schwarzian}Schwarzian Modes for \ads{2}-JT Gravity}

Considering the two-dimensional geometry, since the curvature will determine the metric locally, the \ads{2} spacetime is locally the same, and the \ads{2} spacetime has a full reparametrization symmetry. However, if considering the quantum fluctuations, the IR theory of the \ads{2} spacetime is labeled by the boundary curve, which breaks the reparametrization symmetry \cite{maldacena2016}. Hence, integrating over quantum fluctuations of the metric becomes integrating over the boundary reparametrization modes, which are called the Schwarzian modes. To show how this works, we first consider the \ads{2} spacetime in the Poincar\'{e} patch, the metric is given by
\begin{equation}
    ds^2=\frac{dt^2+dz^2}{z^2}\,,
\end{equation}
with the boundary cutoff curve parametrized by $u\in [0,\beta]$, and at the boundary, the induced metric is given by $h_{uu}=\frac{1}{\varepsilon^2}+o(\varepsilon^{-2})$, where $\varepsilon\rightarrow 0$. Consequently, the boundary cutoff curve is given by $z(u)=\varepsilon t'(u)$. Then, the extrinsic curvature is the Schwarzian derivative, i.e.
\begin{equation}
    K=1+\varepsilon^2\mathrm{Sch}(t,u)+o(\varepsilon^2)\,,
\end{equation}
with
\begin{equation}
    \mathrm{Sch}(t,u)=-\frac{1}{2}\left(\frac{t''}{t'}\right)^2+\left(\frac{t''}{t'}\right)'\,.
\end{equation}
The first term of the extrinsic curvature is a constant that can be canceled by a counterterm expressed as the integral of the boundary dilaton along the cutoff curve.

Integrating out the dilaton field $\Phi$, we obtain
\begin{widetext}
   \begin{equation}
    Z=\int [Dg][D\Phi]e^{-I[g,\Phi]}=\int [Dg]\delta(R+2)e^{-\frac{1}{8\pi G}\int_{\partial M_2}du\Phi_b\mathrm{Sch}(t,u)}\,,
\end{equation}
\end{widetext}
which fixes the scalar curvature of the quantum fluctuating \ads{2} spacetime, as discussed later.

By the chain rule of the Schwarzian derivative, we can define another reparametrization $du_b=\bar{\Phi}_b du/\Phi_b$ such that $\Phi_b$ becomes a constant at the boundary, similar to the discussion in \cite{Maldacena_2003}. Redefining $\bar{\Phi}_b=8\pi GC$, the partition function is totally governed by the Schwarzian derivative:
\begin{equation}
    Z=\int [Dg]\delta(R+2)e^{-C\int_{\partial M_2}du\mathrm{Sch}(t,u)}\, .\label{eq:partition in C}
\end{equation}
Then, we can see that the gravitational effect of the \ads{2} spacetime is governed by the Schwarzian derivative. Hence, considering the conformal gauge, we take the ansatz:
\begin{equation}
    ds^2=-e^{2\omega(x^{\pm})}dx^+dx^-\,,
\end{equation}
where the light-cone coordinates are given by $x^\pm=it\pm z$. The scalar curvature is given by
\begin{equation}
    R=8e^{-2\omega}\partial_+\partial_-\omega.
\end{equation}
Fixing the scalar curvature to be $-2$, we finally obtain the Liouville's equation for $\omega$
\begin{equation}
    4\partial_+\partial_-\omega+e^{2\omega}=0\,,
\end{equation}
whose solution is given by two functions $X^+(x^+)$ and $X^-(x^-)$:
\begin{equation}
    e^{2\omega}=\frac{4\partial_+X^+(x^+)\partial_-X^-(x^-)}{(X^+(x^+)-X^-(x^-))^2}\,.
\end{equation}
Thus, different choices of the function $X^+$ and $X^-$ represent the quantum fluctuations on the \ads{2} spacetime, reparametrizing the light-cone coordinates $x^\pm$ to $X^\pm$. For the \ads{2} spacetime from the dimensional reduction of the RN black hole, the two functions should be the same 
\begin{equation}
    X^+(x)=X^-(x)=F(x)=\frac{\beta}{\pi}\tanh\frac{\pi}{\beta}x\,,
\end{equation}
then the \ads{2} spacetime is given by
\begin{equation}
    ds^2=-\frac{4\pi^2}{\beta^2}\frac{dx^+dx^-}{\sinh^2(\frac{\pi}{\beta}(x^+-x^-))}\,.
\end{equation}
as expected in \eqref{eq:ads2fromrn}. To compactify the coordinates, we may introduce ``complex light-cone coordinates'', such that 
\begin{equation}
    ds^2=-\frac{4\pi^2}{\beta^2}\frac{dudv}{\sin^2(\frac{\pi}{\beta}(u-v))}\,.
\end{equation}
In that case, we can perform a field redefinition, $t=\tan\frac{\pi}{\beta}f(u)$, then the Schwarzian action becomes
\begin{equation}
    I=-C\int_0^\beta du\left(\frac{f'''(u)}{f'(u)}-\frac{3}{2}\frac{f''(u)^2}{f'(u)^2}+\frac{2\pi^2}{\beta^2}f'(u)^2\right)\, .
\end{equation}
For this action, we can introduce an auxiliary field $\psi$, and the action then possesses a supersymmetry, allowing it to be one-loop exact, as shown in \cite{Stanford_2017}. After the introducing the auxiliary $\psi$, the Schwarzian action becomes
\begin{equation}
    I=\frac{C}{2}\int^\beta_0 du\left[(\frac{f''}{f'})^2-\frac{4\pi^2}{\beta^2}(f')^2+\frac{\beta^2}{4\pi^2}\frac{\psi''\psi'}{f'^2}-\psi'\psi\right]\,.\label{action Schwarzian}
\end{equation}

We can apply the perturbation theory to this model. Since $\tau\sim\tau+\beta$, we can always employ mode expansions on $\tau$ and $\psi$:
\begin{align}
    f=&u+g\epsilon(u)=\tau+g\sum_{|n|\geq 2}\frac{\beta}{2\pi}\epsilon_n e^{-i2\pi n u/\beta}\,,\\
    \psi=&g\sum_{|n|\geq 2}\sqrt{\frac{\beta}{2\pi}}\psi_n e^{-i2\pi n u/\beta}\,.
\end{align}

The terms only containing $\sim\epsilon'(u)$ are total derivative terms; hence, they vanish after integration by parts. Therefore, we can get the following action:
\begin{equation}
    I=I^{(0)}+I^{(2)}+gI^{(3)}+g^2I^{(4)}+O(g^3),
\end{equation}
where 
\begin{align}
    I^{(0)}&=-\frac{2\pi^2C}{\beta}\,,\\
    I^{(2)}&=\frac{C}{2}\int d\tau[(\epsilon'')^2-\frac{4\pi^2}{\beta^2}(\epsilon')^2+\frac{\beta^2}{4\pi^2}\psi''\psi'-\psi'\psi]\,,\\
    I^{(3)}&=\frac{C}{2}\int d\tau [-2(\epsilon'')^2\epsilon'-2\frac{\beta^2}{4\pi^2}\psi''\psi'\epsilon']\,,\\
    I^{(4)}&=\frac{C}{2}\int d\tau [3(\epsilon')^2(\epsilon'')^2+3\frac{\beta^2}{4\pi^2}(\epsilon')^2\psi''\psi']\,.
\end{align}

Let us first calculate the action in the momentum space:
\begin{widetext}
        \begin{align}
    I^{(2)}=&\frac{4\pi^2 C}{\beta}\sum_{|n|\geq 2}n^2(n^2-1)\epsilon_{-n}\epsilon_n-iC\sum_{|n|\geq 2}n(n^2-1)\psi_{-n}\psi_n\,,\\
    I^{(3)}=&i\frac{4\pi^2C}{3\beta^2}\sum_{|n|,|m|,|n+m|\geq 2}mn(m+n)(m^2+mn+n^2)\epsilon_{-m-n}\epsilon_m\epsilon_n\notag\\
    &+\frac{C}{3}\sum_{|n|,|m|,|n+m|\geq 2}nm(m+n)(n+2m)\psi_{-m-n}\psi_m\epsilon_n\,,\\
    I^{(4)}=&-\frac{\pi^2 C}{\beta^2}\sum_{\text{all modes} \geq 2}mnp(m+n+p)(m^2+n^2+p^2+mn+mp+np)\epsilon_{-m-n-p}\epsilon_m\epsilon_m\epsilon_p\notag\\
    &+i\frac{3C}{2}\sum_{\text{all modes} \geq 2}mnp(m+n+p)(m+n+2p)\psi_{-m-n-p}\psi_p\epsilon_m\epsilon_n\,.
\end{align}
Here, we have symmetrized bosonic modes and antisymmetrized fermionic modes. Hence, the Feynman rules of the Schwarzian action are given as follows. First, the free propagators are
        \begin{equation}
        \begin{tikzpicture}[baseline={(current bounding box.center)}]
            \coordinate (in) at (-1,0);
            \coordinate (out) at (1,0);
            \draw [graviton] (in) -- (out);
            \node (k) at (-1.3,0) {$\epsilon_{-n}$};
            \node (q) at (1.3,0) {$\epsilon_n$};
        \end{tikzpicture}=\frac{1}{2\pi i}\frac{1}{n^2(n^2-1)},\begin{tikzpicture}[baseline={(current bounding box.center)}]
            \coordinate (in) at (-1,0);
            \coordinate (out) at (1,0);
            \draw [loosely dotted,line width=2pt] (in) -- (out);
            \node (k) at (-1.3,0) {$\psi_{-n}$};
            \node (q) at (1.3,0) {$\psi_n$};
        \end{tikzpicture}=\frac{1}{2\pi i}\frac{1}{n(n^2-1)}\, .
\end{equation}
Performing the inverse Fourier transform, we will get the tree-level propagators of $\epsilon$ as
\begin{equation}
\begin{split}
    \langle \epsilon(u)\epsilon(0)\rangle&=(\frac{\beta^2}{2\pi})^2\sum_{|n|\geq 2}\langle \epsilon_{-n}\epsilon_n\rangle e^{-2\pi inu/\beta}=\frac{1}{2\pi C}(\frac{\beta}{2\pi})^2\sum_{|n|\geq 2}e^{-2\pi inu/\beta}\frac{1}{n^2(n^2-1)}\\
    &=\frac{1}{2\pi C}(\frac{\beta}{2\pi})^3[1+\frac{5}{2}\cos\frac{2\pi u}{\beta}-\mathrm{Li}_2(e^{-2\pi iu/\beta})-\mathrm{Li}_2(e^{2\pi iu/\beta})\\
    &+i(\log(1-e^{-2\pi iu/\beta})-\log(1-e^{2\pi iu/\beta}))\sin\frac{2\pi u}{\beta}]\, ,
\end{split}
\end{equation}
where $\mathrm{Li}_2(x)$ is the polylogarithm function defined as
\begin{equation}
    \mathrm{Li}_{s}(z)=\sum_{n=1}^{\infty}{\frac{z^{n}}{n^{s}}}\,.
\end{equation}
From the properties of the polylogarithm function $\mathrm{Li}_2(\frac{1}{y})+\mathrm{Li}_2(y)=\frac{\pi^2}{3}-\frac{1}{2}\log^2y-i\pi\log y$, and $\log(1-e^{-ix})-\log(1-e^{ix})=-ix+\log(e^{ix}-1)-\log(1-e^{ix})=-ix+i\pi$ we obtain:
\begin{equation}
\begin{split}
    \langle \epsilon(u)\epsilon(v)\rangle&\approx\frac{1}{2\pi C}(\frac{\beta}{2\pi})^3\left[1+\frac{\pi^2}{6}+\frac{5}{2}\cos\frac{2\pi(u-v)}{\beta}-\frac{1}{2}(\frac{2\pi(u-v)}{\beta}-\pi)^2+(\frac{2\pi(u-v)}{\beta}-\pi)\sin\frac{2\pi(u-v)}{\beta}\right]\,.\label{ee-functions tree}
\end{split}
\end{equation}

From the interaction term, we can read off vertices; for instance, the three-point vertices of bosonic and fermionic fields, as well as the four-point vertices of bosonic and fermionic fields, are shown in Fig.~\ref{3pt} and Fig.~\ref{4pt}, respectively.

    \begin{figure}[htb!]
    \centering
    \begin{subfloat}[][$\epsilon\epsilon\epsilon$-vertices]
    {\begin{tikzpicture}[baseline={(current bounding box.center)}]
            \draw [graviton] (-2,0) -- (1,0);
            \draw [graviton] (1,0) -- (3,1.732);
            \draw [graviton] (1,0) -- (3,-1.732);
    \end{tikzpicture}}
    \end{subfloat}\quad\quad
    \begin{subfloat}[][$\epsilon\psi\psi$-vertices]{\begin{tikzpicture}[baseline={(current bounding box.center)}]
            \draw [graviton] (-2,0) -- (1,0);
            \draw [loosely dotted,line width=2pt] (1,0) -- (3,1.732);
            \draw [loosely dotted,line width=2pt] (1,0) -- (3,-1.732);
    \end{tikzpicture}}\end{subfloat}
    \caption{3-point vertices of the Schwarzian action}\label{3pt}
\end{figure}
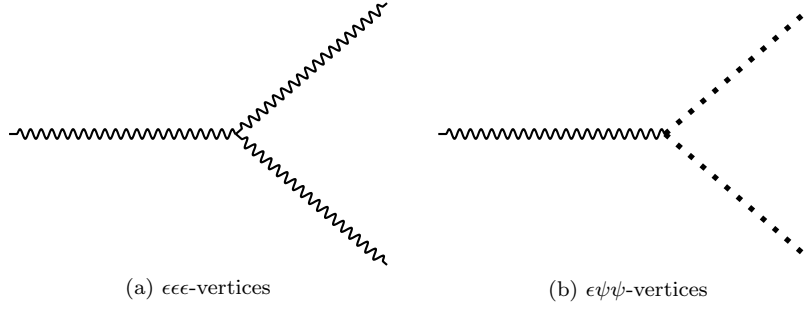
\begin{figure}[htb!]
    \centering
    \begin{subfloat}[][$\epsilon\epsilon\epsilon\epsilon$-vertices]
    {\begin{tikzpicture}[baseline={(current bounding box.center)}]
            \draw [graviton] (-2.2,0) -- (0,0);
            \draw [graviton] (0,0) -- (2.2,0);
            \draw [graviton] (0,0) -- (0,2.2);
            \draw [graviton] (0,0) -- (0,-2.2);
    \end{tikzpicture}}
    \end{subfloat}\quad\quad
    \begin{subfloat}[][$\epsilon\epsilon\psi\psi$-vertices]{\begin{tikzpicture}[baseline={(current bounding box.center)}]
            \draw [graviton] (0,0) -- (0,-2.2);
            \draw [graviton] (-2.2,0) -- (0,0);
            \draw [loosely dotted,line width=2pt] (0,0) -- (2.2,0);
            \draw [loosely dotted,line width=2pt] (0,0) -- (0,2.2);
    \end{tikzpicture}}\end{subfloat}
    \caption{4-point vertices of the Schwarzian action}\label{4pt}
\end{figure}
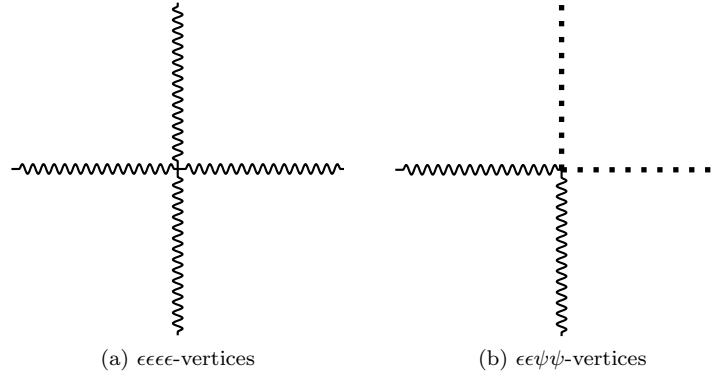

Calculating the one-loop two-point function of the reparametrization, \cite{Qi_2019} showed that after considering the fermionic loop, the two-point functions may be written in the momentum space as
    \begin{equation}
    \langle\epsilon_{-n}\epsilon_n\rangle_{\mathrm{cubic}}=\begin{cases}
    \frac{g^2}{8\pi^2}\frac{-112+104n^2+179n^4-111n^6+12n^8}{n^2(n^2-4)^2(n^2-1)^2},\quad n\neq 2\,,\\
    -\frac{g^2}{96}+\frac{1465g^2}{6912\pi^2},\quad n=2\,,
    \end{cases}
\end{equation}
\begin{equation}
    \langle \epsilon_{-n}\epsilon_n\rangle_{\mathrm{quartic}}=-\frac{9g^2}{8\pi^2}\frac{1}{(n^2-1)^2}\,,
\end{equation}
\end{widetext}
which remain finite after the inverse Fourier transform.

In this case, we can calculate the quantum corrections to the quantities in the \ads{2} spacetime. For the \ads{2} metric, we can expand the metric in terms of the Schwarzian modes $\epsilon$'s. Up to the quadratic order in $\epsilon$, the metric is
\begin{widetext}
    \begin{equation}
    \begin{split}
        ds^2 = & -\frac{4\pi^2}{\beta^2}\frac{f'(x^+)f'(x^-)}{\sin^2\left(\frac{\pi}{\beta}(f(x^+)-f(x^-))\right)}dx^+dx^-\\
        = & -\frac{4\pi^2}{\beta^2}\frac{(1+\epsilon'(x^+))(1+\epsilon'(x^-))}{\sin^2\left(\frac{\pi}{\beta}((x^+-x^-)+(\epsilon(x^+)-\epsilon(x^-))\right)}dx^+dx^-\\
        = & -\frac{4\pi^2}{\beta^2}\frac{dx^+dx^-}{\sin^2\left(\frac{\pi}{\beta}(x^+-x^-)\right)}\Bigg\{(1+\epsilon'(x^+)+\epsilon'(x^-)-\frac{2\pi}{\beta} \cot \left(\frac{\pi(x^+-x^-)}{\beta}\right) (\epsilon(x^+)-\epsilon(x^-))\\
        & \qquad +\frac{1}{2\beta^2}\frac{1}{\sin^2\frac{\pi}{\beta}(x^+-x^-)} \Bigg[4\pi^2(\epsilon(x^+)-\epsilon(x^-))^2+\beta^2\epsilon'(x^+)\epsilon'(x^-)\\
        & \qquad\qquad\qquad\qquad\qquad\qquad - 2\pi\beta \sin \left(\frac{2\pi(x^+-x^-)}{\beta}\right) (\epsilon'(x^+)+\epsilon'(x^-))(\epsilon(x^+)-\epsilon(x^-))\\
        & \qquad\qquad\qquad\qquad\qquad\qquad + \cos \left(\frac{2\pi(x^+-x^-)}{\beta}\right) (2\pi^2(\epsilon(x^+)-\epsilon(x^-))^2-\beta^2\epsilon'(x^+)\epsilon'(x^-))\Bigg]\Bigg\}\,.
    \end{split}\label{metric as ee}
\end{equation}
\end{widetext}
Similar to the quantization law in quantum mechanics, we need to represent a classical operator $AB$ as $\frac{1}{2}(AB+BA)$, which makes the quantum metric operator symmetric in $x^+$ and $x^-$.

\subsection{\label{subse:quantum corrected eom}Quantum-Corrected Scalar Equation of Motion}

In this section, we will derive the quantum correction to the scalar field equation in the RN black hole.

As derived in Sec.~\ref{subsec:Rn to ads2}, the near-horizon region of the near-extremal RN black hole is \ads{2}$\times S^2$. From the dimensional reduction, we will get the JT gravity, which plays the role of the effective action for gravity. Even though the \ads{2} region is located in the near-horizon region, we can still receive the correction for the far-away region. First, the near-horizon region and the far-away region share the same boundary, i.e., the overlapping region. Hence, the equation of motion could be matched in the overlapping region, resulting in the far-away region having the same form of fluctuations as the Schwarzian modes. In that case, the expression of the quantum correction could be carried to the far-away region. Secondly, since the boundary conditions can be matched if they apply to the same equation, if the fields in the RN black hole background are affected by the quantum fluctuations in the near-horizon region, it should also carry information to the far-away region. In addition, it is not hard to find that the correction to the metric vanishes if we take the near-boundary limit $2z=x^+-x^-\rightarrow 0$. Hence, we can calculate the quantum correction in the near-horizon \ads{2} region, and then extend it to the far-away region, which gives us the quantum fluctuations in the whole RN black hole. In the discussion of Sec.~\ref{subsec:scalar qnms}, we will show that for the scalar quasi-normal modes, the only difference between the classical and the quantum-corrected RN black holes is the effective potential, while the boundary conditions on both sides will not be changed, neither does the light ring, the maximum point of the effective potential. Therefore, the global property of the scalar quasi-normal mode equation remains unchanged.

In that case, the Schwinger-Dyson equation \cite{Schwinger, Dyson} for the matter field $\Psi$ in the RN black hole background should be calculated in the following steps, 
    \begin{equation}
    \begin{split}
        \frac{1}{Z}\frac{\delta Z}{\delta\Psi}=&\frac{1}{Z}\int[d\Psi][dg_{4d}]\frac{\delta I_{\mathrm{mat}}}{\delta\Psi}e^{-I_{\mathrm{gra}}[g]-I_{\mathrm{mat}}[g,\Psi]}\\
        =&\frac{1}{Z}\int[d\Psi][d\Phi][dg_{2d}]\frac{\delta I_{\mathrm{mat}}}{\delta\Psi}e^{-I_{\mathrm{gra}}[g,\Phi]-I_{\mathrm{mat}}[g,\Psi]}\\
        =&\frac{1}{Z}\int [d\Psi][dg_{2d}]\delta(R+2)\frac{\delta I_{\mathrm{mat}}}{\delta\Psi}e^{-I_{\mathrm{gra}}[g_{2d},\Phi]-I_{\mathrm{mat}}[g_{2d},\Psi]}\\
        =&\frac{1}{Z}\int [d\Psi][d\epsilon]\frac{\delta I_{\mathrm{mat}}}{\delta\Psi}e^{-I_{\mathrm{Schwarzian}}[\epsilon]-I_{\mathrm{mat}}[g_{2d}(\epsilon),\Psi]}\\
        = & \Big\langle\frac{\delta I_{\mathrm{matter}}}{\delta\Psi}\Big\rangle\,,
    \end{split}
\end{equation}
which means that we should first write the metric as a function of the Schwarzian modes, like in Eq.~\eqref{metric as ee}. Then, we can regard it as a given metric and derive the equation of motion for the matter fields. Finally, we should average the matter equation of motion within the path integral over the Schwarzian modes, which will eventually provide us with the quantum-corrected equation of motion. In principle, this equation is not equivalent to the classical equation of motion for a matter field in the quantum-corrected metric. However, these two treatments coincide at the leading order in quantum correction for a neutral scalar field, in the same spirit as the Ehrenfest theorem in quantum mechanics \cite{Ehrenfest1927}.

For the free massless real scalar field, we can take the ansatz as:
\begin{equation}
    \Psi(r)=\sum_{l,m,\omega}\frac{R(r)}{r}e^{-i\omega t}Y_{lm}(\theta,\varphi)\,.
\end{equation}
Then, for any spherically-symmetric metric, we can write it as $ds^2=-F(r)dt^2+G(r)dr^2+r^2d\theta^2+r^2\sin^2\theta d\varphi^2$. Hence, the equation of motion of the scalar field becomes:
\begin{align}
        \frac{1}{\sqrt{F(r)G(r)}r^2}&\left(-\partial_tr^2\sqrt{\frac{G(r)}{F(r)}}\partial_t+\partial_rr^2\sqrt{\frac{F(r)}{G(r)}}\partial_r\right)\Psi \nonumber \\
        -&\frac{l(l+1)}{r^2}\Psi=0\,.
\end{align}
For the metric of the form \eqref{metric as ee}, we see that the metric components have the form:
\begin{equation}
    \begin{split}
        F(r) = & \left(1-\frac{2M}{r}+\frac{Q^2}{r^2}\right) A_{\text{cor}}=f(r)A_{\text{cor}}\,,\\
        G(r) = & \frac{A_{\text{cor}}}{1-\frac{2M}{r}+\frac{Q^2}{r^2}}=\frac{A_{\text{cor}}}{f(r)}\, .
    \end{split}
\end{equation}
Hence, the scalar field equation in the radial direction becomes:
\begin{equation}
    \begin{split}
    0=\frac{1}{r}\frac{d}{dr}(f(r)\frac{d}{dr}R_{lm}(r))\\
    + \left(\frac{1}{rf(r)}\omega^2-A_{\mathrm{cor}}\frac{l(l+1)}{r^3}-\frac{f'(r)}{r^2}\right) R_{lm}(r)\label{eqn: corrected r-direction}\,,
    \end{split}
\end{equation}
where the correction factor is given as
\begin{widetext}
    \begin{equation}
    \begin{split}
        A_{\text{cor}} = & \sin^2\left(\frac{\pi}{\beta}(x^+-x^-)\right)\times\left[\frac{(1+\epsilon'(x^+))(1+\epsilon'(x^-))}{\sin^2\left(\frac{\pi}{\beta}((x^+-x^-)+(\epsilon(x^+)-\epsilon(x^-))\right)}\right]\\
        =\, & 1+\epsilon'(x^+)+\epsilon'(x^-)-\frac{2\pi}{\beta} \cot\left(\frac{\pi(x^+-x^-)}{\beta}\right) (\epsilon(x^+)-\epsilon(x^-))\\
        {} & +\frac{1}{2\beta^2}\frac{1}{\sin^2\frac{\pi}{\beta}(x^+-x^-)}\Bigg[4\pi^2(\epsilon(x^+)-\epsilon(x^-))^2+\beta^2\epsilon'(x^+)\epsilon'(x^-)\\
        {} & -2\pi\beta \sin\left(\frac{2\pi(x^+-x^-)}{\beta}\right)(\epsilon'(x^+)+\epsilon'(x^-))(\epsilon(x^+)-\epsilon(x^-))\\
        {} & + \cos\left(\frac{2\pi(x^+-x^-)}{\beta}\right) (2\pi^2(\epsilon(x^+)-\epsilon(x^-))^2-\beta^2\epsilon'(x^+)\epsilon'(x^-))\Bigg]\,.    \end{split}\label{cor-factor as ee}
\end{equation}
\end{widetext}
Here, $x^+$ and $x^-$ are the light-cone coordinates for the reduced \ads{2} spacetime. Putting this equation into the path integral, we obtain the quantum-averaged correction factor $\langle A_{\mathrm{cor}}\rangle$ in the RN black hole coordinates as:
\begin{widetext}
    \begin{align}
     \langle A_{\mathrm{cor}}\rangle = 1+\frac{\beta}{32\pi^4C}(r-r_+)(r-r_-) & \Bigg[6\pi^2\ln^2\frac{r-r_-}{r-r_+} + (6\pi^2+4\pi^2\ln^2\frac{r-r_-}{r-r_+})\frac{(2\pi/\beta)^2}{(r-r_+)(r-r_-)} \nonumber\\
    & -6\beta\pi\ln\frac{r-r_-}{r-r_+}\frac{(2\pi/\beta)(r-r_h)}{(r-r_+)(r-r_-)}\Bigg]\,.
    \end{align}
\end{widetext}

\newpage

\newpage
Finally, we obtain the radial equation of motion for the scalar field after quantum correction
\begin{equation}
    \begin{split}
    0=\frac{1}{r}\frac{d}{dr}(f(r)\frac{d}{dr}R_{lm}(r))\\
    + \left(\frac{1}{rf(r)}\omega^2-\langle A_{\mathrm{cor}}\rangle\frac{l(l+1)}{r^3}-\frac{f'(r)}{r^2}\right) R_{lm}(r)\label{eqn: corrected r-direction}\,.
    \end{split}
\end{equation}

\subsection{Evaluation of the Boundary Dilaton\label{subsec: boundary dilaton}}
For the \ads{2} JT gravity, assuming the boundary dilaton field is a constant, it can be viewed as a free parameter. However, this is not right for the \ads{2} near-horizon region of the RN black hole. In \cite{Iliesiu_2021}, considering the Schwarzian action correction, the entropy of the RN black hole is given by
\begin{equation}
    S=S_0+{4\pi^2\bar{\Phi}_b}T+\frac{3}{2}\log({\bar{\Phi}_bT})+O({T^2})\,,\label{entropy}
\end{equation}
while from the classical black hole thermodynamics, we obtain the semiclassical entropy as
\begin{equation}
    S=\frac{A_{\mathrm{H}}}{4}=\pi r_+^2=\pi M^2+8\pi^2 M^3T+O(T^2)\,.
\end{equation}
In the high-temperature limit, the logarithmic term is subleading, and hence, comparing the terms linear in temperature, which is the capacity of the black hole, it can be deduced that the boundary dilaton field is given by
\begin{equation}
    \bar{\Phi}_b={2M^3}\,.
\end{equation}
Since the boundary dilaton $\bar{\Phi}_b$ is proportional to the coupling $C$, the strength of the quantum correction is determined by the mass of the black hole:
\begin{equation}
    C\sim M^3\,.\label{eq: evaluation of C}
\end{equation}
From dimensional analysis, we can determine the coupling $C$ in SI units,
\begin{equation}
\begin{split}
    CT=&16\pi G\frac{M^3}{m_{\mathrm{pl}}^3c^4l_{\mathrm{pl}}}{k_BT}\\
    =&16\pi G\frac{M^3}{(\hbar c/G)^{3/2}c^4(\hbar G/c^3)^{1/2}}k_BT\\
    =&\frac{16\pi G^2k_B}{\hbar^2c^4}M^3T\,,
\end{split}
\end{equation}
where the Hawking temperature of the RN black hole is
\begin{equation}
    T=\frac{\hbar c^3}{8\pi G k_B M}\left(\frac{\sqrt{1-\frac{Q^2}{4\pi\epsilon_0 G M^2}}}{\left(1+\sqrt{1-\frac{Q^2}{4\pi\epsilon_0 G^2M^2}}\right)^2}\right)\,.
\end{equation}
Therefore, 
\begin{equation}
    CT=\frac{2G}{\hbar c}M^2\left(\frac{\sqrt{1-\frac{Q^2}{4\pi\epsilon_0 G M^2}}}{\left(1+\sqrt{1-\frac{Q^2}{4\pi\epsilon_0 G^2M^2}}\right)^2}\right)\,.
\end{equation}
This expression shows that when the mass $M$ is sufficiently small while the electric charge $Q$ is large, $CT$ becomes small. In this regime, higher-order quantum corrections should be taken into account.

Meanwhile, the expression of the quantum-corrected entropy gives us the effective range for our discussion. The lower bound of the entropy is $-\infty$, and in our discussion, the temperature of the RN black hole could not be so low that the entropy is negative. Therefore, our discussion has three requirements: (1) It is a near-extremal RN black hole, i.e., the temperature $T$ should be smaller than that of a Schwarzschild black hole with the same mass, i.e., $T\ll (8\pi M)^{-1}$; (2) The perturbative expansion is suitable, i.e., $\beta/C\ll 1$; (3) The quantum-corrected entropy should be positive, i.e., the entropy given in \eqref{entropy} is positive.

\section{\label{sec:Scalar QNMs}Calculation of Scalar QNMs}

\subsection{Basic Discussion on Scalar Quasi-Normal Modes}\label{subsec:scalar qnms}

\begin{figure}[htb!]
\includegraphics[width=9cm]{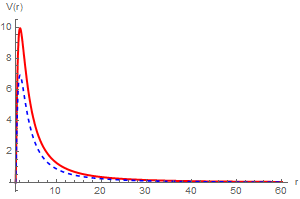}
\caption{\label{fig:potential} The effective potential. The blue line and the red line represent the classical and the quantum corrected effective potentials for the scalar field, respectively, in the RN black hole background.}
\end{figure}

For the scalar field equation, using the tortoise coordinates $x=\int dr/f(r)$. Then, the equation of motion for the scalar field becomes:
\begin{equation}
    -\frac{d^2}{dx^2}R_{lm}+(V(x)-\omega^2)R_{lm}=0\,,
\end{equation}
where the effective potential is given by
\begin{equation}
    V(x)\equiv V(r(x))=f(r)\left(\frac{l(l+1)}{r^2}\langle A_{\mathrm{cor}}\rangle+\frac{f'(r)}{r}\right).\label{eq: potential}
\end{equation}
The corrected potential and the original RN case are shown in Fig.~\ref{fig:potential}.

The effective potential after correction has a similar property to the classical case. First, the effective potential vanishes at the horizon $r=r_+$ and at infinity $r=\infty$. Additionally, the effective potential has only one peak, and numerical calculations reveal that the maximum position is nearly identical to the original RN case. Hence, this equation is an eigenvalue problem, determining the discrete spectrum of $\omega$. We have the boundary conditions at the horizon and infinity as
\begin{equation}
    \begin{split}
        R\rightarrow e^{-i\omega x}&\,,\quad x\rightarrow -\infty\,,\\
        R\rightarrow e^{i\omega x}&\,,\quad x\rightarrow \infty,
    \end{split}
\end{equation}
which impose the ingoing boundary condition at the horizon and the outgoing boundary condition at the future infinity.

Several methods have been developed for determining QNMs, including the Wentzel–Kramers–Brillouin (WKB) method \cite{Iyer1987I}, continuous fraction method \cite{Leaver:1985ax}, Prony method \cite{Berti_2007}, spectral method \cite{Jansen:2017oag}, and so forth. Among these methods, the WKB method has been developed in numerous papers, ranging from the 1st order to the 13th order. However, it is important to note that the higher-order WKB method does not necessarily imply higher accuracy. In later calculations, we find that the 3rd- and 4th-order WKB results are the most stable ones, having very small errors compared to the results given by the Prony method (see App.~\ref{app:diff order}). Therefore, we will use the 3rd-order WKB method and the Prony method to obtain the QNMs in our work. In the following two subsections, we will introduce these two methods.

\subsection{WKB Method\label{sec:WKB}}
For the Schr\"{o}dinger-like equation in tortoise coordinate $x$, 
\begin{equation}
\frac{d^2 R}{dx^2} + \left[\omega^2 - V(x) \right] R = 0\,,
\end{equation}
there is a maximum point at $x=x_0$
\begin{equation}
\left. \frac{dV}{dx} \right|_{x = x_0} = 0\,, \quad V_0 = V(x_0)\,.
\end{equation}
Hence, the effective potential can be approximated by the power series
\begin{equation}
    V(x)=V_0+\frac{1}{2}V''(x_0)(x-x_0)^2+\ldots\,.
\end{equation}

Then, the quasi-normal modes can be derived from perturbation theory,
\begin{equation}
i \frac{\omega^2 - V_0}{\sqrt{-2 V''(x_0)}} = n + \frac{1}{2}+\sum_{j}\Lambda_j\,,\label{eq:WKB}
\end{equation}
where $n$ is an integer named the overtone number of the quasi-normal modes, and $\Lambda_i$ is the $i$-th order correction to the quasi-normal modes, which can be found in \cite{Iyer1987I, IyerII, Konoplya_2019}.

\subsection{Finite difference and Prony Method}

Besides the WKB approximation method, we can also obtain the quasi-normal modes from the time evolution of the fields, which was first developed in \cite{Gundlach_1994}. For the time-dependent Schr\"{o}dinger-like equation, we can write it as
\begin{eqnarray}
    -\frac{\partial^2\Psi(t,x)}{\partial t^2}+\frac{\partial^2\Psi(t,x)}{\partial x^2}=V(x)\Psi(t,x)\,.
\end{eqnarray}
Introducing the light-cone coordinates $u=t+x$ and $v=t-x$, we obtain 
\begin{eqnarray}
    \partial_u\partial_v\Psi(u,v)=-V\Psi(u,v)\,.
\end{eqnarray}
Suppose there is a lattice such that $\Psi_{ij}=\Psi(u_0+ih,v_0+jh)$. Rewriting the derivative as a finite difference
\begin{equation}
\begin{split}
    \frac{d\Psi_{i,j}}{du}=\frac{\Psi_{i+1,j}-\Psi_{i-1,j}}{2h}\,,\\
    \frac{d\Psi_{i,j}}{dv}=\frac{\Psi_{i,j+1}-\Psi_{i,j-1}}{2h}\,,\label{field eom}
\end{split}
\end{equation}
redefining $\Psi_{i+1,j+1}\rightarrow \Psi_N,\Psi_{i-1,j-1}\rightarrow \Psi_S,\Psi_{i+1,j-1}\rightarrow \Psi_E,\Psi_{i-1,j+1}\rightarrow \Psi_W$, and using the central rule as
\begin{equation}
    \Psi_{i,j}=\frac{\Psi_W+\Psi_E}{2}=\frac{\Psi_{i+1,j-1}+\Psi_{i-1,j+1}}{2}\,,
\end{equation}
we obtain a finite difference equation
\begin{equation}
    \Psi_{N}=\Psi_W+\Psi_E-\Psi_S-4h^2V_{ij}\frac{\Psi_W+\Psi_E}{2}\,.
\end{equation}
Given any initial data, for instance, the Gaussian wave packet $\Psi(u,0)=0,~\Psi(0,v)=\exp(-(v-v_0)^2/\sigma^2)$, we can get $\Psi$ for any $u$ and $v$. For example, the following two figures show the time evolution of the scalar with or without quantum corrections.
\begin{figure}[htb!]
\includegraphics[width=9cm]{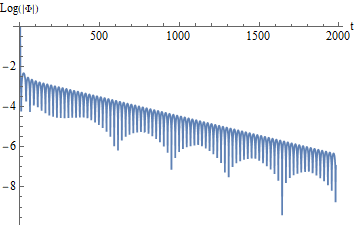}
\caption{\label{fig:evo without cor} The time evolution of the scalar field in the RN black hole background without quantum correction}
\end{figure}

\begin{figure}[htb!]
\includegraphics[width=9cm]{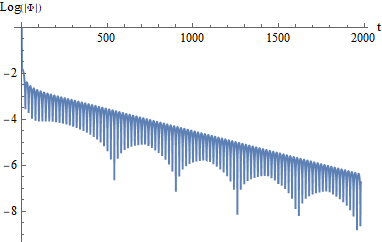}
\caption{\label{fig:evo with cor} The time evolution of the scalar field in the RN black hole background with quantum correction}
\end{figure}

We can transform the light-cone coordinates back to the spacetime coordinates. In that case, the time evolution can be written as:
\begin{equation}
    x_j=\Psi(t_0+jh)=\sum_{n=1}^p A_n z^j_n\,,
\end{equation}
where $A_n=C_ne^{-i\omega_n t_0}$ is the initial data of the time evolution. Hence, we can obtain the quasi-normal modes through $z_n=\exp(-i\omega_n h)$. Consider a polynomial
\begin{equation}
    P(z)=\prod_{n=1}^p(z-z_n)\,,
\end{equation}
which can be expanded to a degree $p$ polynomial
\begin{equation}
    P(z)=\sum_{n=0}^p\alpha_n z^{p-n}\,,\label{expansion}
\end{equation}
where the zeros of this polynomial are given by the $z_n$'s.

It is not hard to show that:
\begin{equation}
    \begin{split}
        \sum_{n=0}^p\alpha_n x_{m-n}&=\sum_{n=0}^p\alpha_n\sum_{k=1}^pA_kz_k^{m-n}\\
        &=\sum_{k=1}^pA_kz_k^{m-p}\sum_{n=0}^p\alpha_n z_k^{m-n}\\
        &=\sum_{k=1}^p A_k z_k^{m-n} P(z=z_k)=0\,,
    \end{split}
\end{equation}
where we have used Eq.~\eqref{expansion}. Hence, there is a linear equation about $x_m(m\geq p)$
\begin{equation}
    x_m=-\sum_{n=1}^p\alpha_nx_{m-n}\,.
\end{equation}
Solving this equation, we finally arrive at the quasi-normal modes in which the fundamental mode has the largest amplitudes.

\section{\label{sec:results}Results of quantum-corrected scalar quasi-normal modes}

\begin{figure}[htb!]
    \begin{center}
    \includegraphics[width=0.95\linewidth]{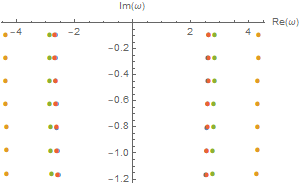}
    \end{center}
    \caption{Quasi-normal modes of the scalar field with fixed mass and charge-mass ratio $q=0.9999$, quantum number $l=10$, and different values of $CT$. The values of $CT=10^4,\,10^5,\,10^6$ are denoted by the yellow, green, and red dots. The red dots almost coincide with the classical results given by the blue dots.}
    \label{fig:frequency}
\end{figure}

\begin{figure}[htb!]
\includegraphics[width=9cm]{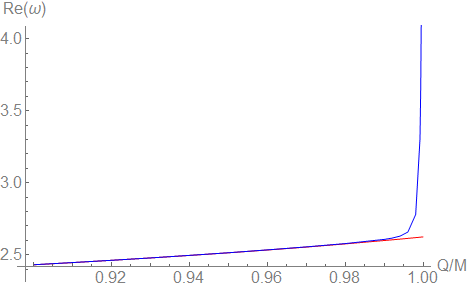}
\caption{\label{fig:real ratio} The real part of quasi-normal modes as a function of $Q/M$ with a given mass $M=1~\mathrm{kg}$. The red and the blue curves denote the classical and the quantum-corrected results, respectively.}
\end{figure}

\begin{figure}[htb!]
\includegraphics[width=9cm]{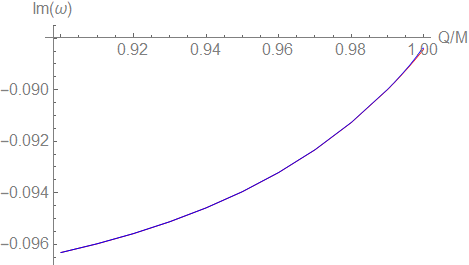}
\caption{\label{fig:Imaginary ratio} The imaginary part of quasi-normal modes as a function of $Q/M$ with the given mass $M=1~\mathrm{kg}$. The red and blue curves denote the classical and the quantum-corrected results, respectively.}
\end{figure}

To see the effects of the mass and the charge-mass ratio more clearly, we fix one parameter and change the other parameter, and then we obtain the results shown in Fig.~\ref{fig:frequency}, \ref{fig:real ratio}, \ref{fig:Imaginary ratio} and for detailed data, see App.~\ref{app:datas}.


First, the quasi-normal modes are modified by the quantum correction to the equation of motion. In particular, the real parts of the quasi-normal modes have much larger quantum corrections than the imaginary parts.

When the black hole becomes more extremal, in other words, the temperature is lower, the quantum correction becomes stronger. From dimensional analysis, $CT$ is proportional to the $\frac{\delta}{M}\times M^2\times 10^5$ with the mass unit kg. In that case, when the mass is about $10^{-3}$kg, the quantum correction is so strong that the perturbation theory breaks down. In that case, quantum gravity is needed when analyzing this phenomenon.

\section{Discussion and Outlook}\label{sec:discussion}

In this paper, we introduce the quantum gravity correction to the scalar field in the RN black hole background and discuss its quasi-normal modes. With the 2d JT gravity/1d Schwarzian theory correspondence, the effective equation of motion for the scalar field can be calculated by the path integral over all reparametrization modes.

In comparison to the original scalar quasi-normal modes in the RN black hole background, the only difference is the effective potential in the quantum correction equation. It can be seen that the quasi-normal modes differ from the uncorrected case. Compared with the classical modes, the quantum correction should be large if the temperature is low or the mass is small. For the quasi-normal modes, the real part is very sensitive to the quantum correction, while the imaginary part is not. In our discussion, if $CT<1$, the quantum correction becomes very large, but it should be noted that here we use the perturbative expansion of the Schwarzian modes, and hence, the $CT<1$ results are not valid in this limit. The exact quantum correction of the RN black hole should be carried out in non-perturbative methods later, which can be one of the further directions to investigate.

From numerical analysis, for the small mass near-extremal RN black hole, the quasi-normal modes may be completely different from the classical solutions. In that case, our calculation may provide a new perspective on primordial black holes, as their masses could be very small.

In addition, other perturbation fields may also be defined on the RN black hole background, and further calculations can be achieved using our methods, such as spinor fields, vector fields, and tensor fields. The quantum corrections of the RN black hole may cause many other observable effects of physical interest.

\section*{Acknowledgments}
We would like to thank Ying-Jian Chen, Xiao-Long Liu, Ren\'{e} Meyer, Run-Qiu Yang, Zhenbin Yang, Amos Yarom, and Cong-Yuan Yue for many helpful discussions. J.N. was partially supported by the NSFC under grants No.~12375067, No.~12147103, and No.~12247103. C.Y.S. was partially supported by the NSFC under grants No.~12447182. This work is partially supported by the National Natural Science Foundation of China under grants No.~12375058, No.~12361141825, and No.~12035016, as well as the National Key Research and Development Program of China with Grant No.~2021YFC2203001.

\appendix
\section{More Details of the Results}

\subsection{\label{app:diff order}Different Orders of WKB}

As stated in Sec.~\ref{sec:WKB}, the higher-order WKB method does not necessarily mean more accuracy. In this appendix, we use the 1-13 order WKB method to evaluate the scalar quasi-normal modes with $M=10~\mathrm{kg}\, , q=0.9999$, and $l=10$. It can be seen that for the high-order WKB method, the quasi-normal modes become very unstable; hence, we should not trust these results.
\begin{center}
        \begin{table}[htb!]
        \centering
        \begin{tabular}{|c|c|c|}
        \hline
            Order & Corrected & Uncorrected  \\
            \hline
            1 & 3.15307-0.0884334\I & 2.62920-0.0884558\I \\
            \hline
            2 & 3.14966-0.0885292\I & 2.62511-0.0885936\I \\
            \hline
            3 & 3.14966-0.0884130\I & 2.62510-0.0884262\I \\
            \hline
            4 & 3.14987-0.0884071\I & 2.62511-0.0884260\I \\
            \hline
            5 & 3.98150-2.43690\I & 2.62511-0.0884263\I \\
            \hline
            6 & 0.0404317-239.973\I & 2.62511-0.0884262\I \\
            \hline
            7 & 9931.86+9934.76\I & 2.62511-0.0884288\I \\
            \hline
            8 & 184.073+536040.2\I & 2.62533-0.0884215\I \\
            \hline
            9 & 1.05809$\times 10^7$ + 1.05945$\times 10^7$\I & 2.62546-0.0921458\I \\
            \hline
            10 & 35666.7+3.14296$\times 10^9$\I & 2.74253-0.0882124\I \\
            \hline
            11 & 1.79356$\times 10^{11}$+1.79383$\times 10^{11}$\I & 21.7580-21.5846\I \\
            \hline
            12 & 4.47426$\times 10^{13}$+7.19078$\times 10^8$\I & 136.716-3.43513\I \\
            \hline
            13 & 1.90853$\times 10^{15}$+1.90800$\times 10^{15}$\I & 1780.12-1774.87\I \\
            \hline
        \end{tabular}
        \caption{Different-order results}
        \label{tab:WKB order}
    \end{table}
\end{center}

From this table, for the classical RN black hole, if we use the WKB method with orders higher than 10, the results also become very large, which implies the failure of the high-order WKB method. For the quantum-corrected case, the failure order becomes lower, even the five-order WKB method causes errors. In that case, to ensure the reliability of the results, we always employ the three-order WKB method.

\subsection{\label{app:datas}The Comparison of the WKB and the Prony Methods}

To show quantum correction on the quasi-normal modes, we take the quantum numbers $l=1$ and $10$, $n=0$, and use both methods in the previous section to calculate the quasi-normal modes with different masses and charge-mass ratios.

\begin{widetext}

    \begin{table}[htb!]
    \centering
    \begin{tabular}{|c|c|c|c|c|}
        \hline
Quantum number $l$ & Methods & Corrected or not & quasi-normal modes $\omega/M$ & corrections\\ \hline
\multirow{4}{*}{$l_1=1$} & WKB & NO & 0.350922-0.0971917\I  & \multirow{2}{*}{Almost 0}\\ \cline{2-4}
 & WKB & YES &  0.350922-0.0971919\I  & \\ \cline{2-5}
 & Prony & NO  &  0.352210-0.0967679\I & \multirow{2}{*}{1.4$\times 10^{-5}$+1.96$\times 10^{-5}$\I} \\ \cline{2-4}
 & Prony & YES & 0.352224-0.0967875\I & \\ \hline
 \multirow{4}{*}{$l_2=10$} & WKB & NO & 2.43173-0.0963221\I &\multirow{2}{*}{Almost 0} \\ \cline{2-4}
 & WKB & YES & 2.43173-0.0963221\I & \\ \cline{2-5}
 & Prony & NO & 2.43211-0.0962772\I &\multirow{2}{*}{Almost 0}\\ \cline{2-4}
 & Prony & YES & 2.43211-0.0962772\I & \\ \hline
    \end{tabular}
    \caption{Result 1: $M=1000~\mathrm{kg},q=0.9$}
\end{table}

\begin{table}[htb!]
    \centering
    \begin{tabular}{|c|c|c|c|c|}
        \hline
Quantum number $l$ & Methods & Corrected or not & quasi-normal modes $\omega/M$ & corrections\\
\hline
\multirow{4}{*}{$l_1=1$} & \multirow{2}{*}{WKB} & NO & 0.375416-0.0895061\I
 & \multirow{2}{*}{5$\times 10^{-6}$-1.91$\times 10^{-5}$\I}  \\ \cline{3-4}
 &  & YES & 0.375421-0.0895252\I & \\ \cline{2-5}
 & \multirow{2}{*}{Prony} & NO  & 0.377334-0.0894164\I & \multirow{2}{*}{1.6$\times 10^{-5}$-7$\times 10^{-7}$\I}  \\ \cline{3-4}
 &  & YES & 0.377350-0.0894171\I & \\ \hline
 \multirow{4}{*}{$l_2=10$} & \multirow{2}{*}{WKB} & NO & 2.62276-0.0885827\I & \multirow{2}{*}{Almost 0}  \\ \cline{3-4}
 &  & YES & 2.62276-0.0885826\I & \\ \cline{2-5}
 & \multirow{2}{*}{Prony} & NO & 2.62559-0.0883785\I & \multirow{2}{*}{Almost 0}  \\ \cline{3-4}
 &  & YES & 2.62564-0.0883784\I & \\ \hline
    \end{tabular}
    \caption{Result 2: $M=1000~\mathrm{kg},q=0.9999$}
\end{table}

\begin{table}[htb!]
    \centering
    \begin{tabular}{|c|c|c|c|c|}
        \hline
Quantum number $l$ & Methods & Corrected or not & quasi-normal modes $\omega/M$ & corrections\\
\hline
\multirow{4}{*}{$l_1=1$} & WKB & NO & 0.350922-0.0971917\I
 &\multirow{2}{*}{Almost 0} \\ \cline{2-4}
 & WKB & YES & 0.350922-0.0971915\I
 &\\ \cline{2-5}
 & Prony & NO  & 0.352224-0.0967875\I &\multirow{2}{*}{Almost 0} \\ \cline{2-4}
 & Prony & YES & 0.352220-0.0967915\I   & \\ \hline
 \multirow{4}{*}{$l_2=10$} & WKB & NO & 2.43173-0.0963221\I
 &\multirow{2}{*}{Almost 0} \\ \cline{2-4}
 & WKB & YES & 2.43173-0.0963221\I
 & \\ \cline{2-5}
 & Prony & NO & 2.43211-0.0962772\I &\multirow{2}{*}{Almost 0} \\ \cline{2-4}
 & Prony & YES & 2.43211-0.0962772\I   &  \\ \hline
    \end{tabular}
    \caption{Result 3: $M=100~\mathrm{kg},q=0.9$}
\end{table}

\begin{table}[htb!]
    \centering
    \begin{tabular}{|c|c|c|c|c|}
        \hline
Quantum number $l$ & Methods & Corrected or not & quasi-normal modes $\omega/M$ & corrections\\
\hline
\multirow{4}{*}{$l_1=1$} & WKB & NO & 0.375675-0.0893758\I
 &\multirow{2}{*}{7.36$\times 10^{-4}$+1.07$\times 10^{-5}$\I}\\ \cline{2-4}
 & WKB & YES & 0.376411-0.0893651\I
 & \\ \cline{2-5}
 & Prony & NO  & 0.377334-0.0894164\I &\multirow{2}{*}{7.43$\times 10^{-4}$-7.3$\times 10^{-6}$\I} \\ \cline{2-4}
 & Prony & YES & 0.378077-0.0894237\I & \\ \hline
 \multirow{4}{*}{$l_2=10$} & WKB & NO & 2.62510-0.0884262\I
 &\multirow{2}{*}{5.77$\times 10^{-3}$+1.5$\times 10^{-6}$\I}\\ \cline{2-4}
 & WKB & YES & 2.63087-0.0884257\I
 &\\ \cline{2-5}
 & Prony & NO & 2.62558-0.0883785\I  & \multirow{2}{*}{5.77$\times 10^{-3}$+5$\times 10^{-7}$\I}\\ \cline{2-4}
 & Prony & YES & 2.63135-0.0883780\I & \\ \hline
    \end{tabular}
    \caption{Results 4: $M=100~\mathrm{kg},q=0.9999$}
\end{table}

\begin{table}[htb!]
    \centering
    \begin{tabular}{|c|c|c|c|c|}
        \hline
Quantum number $l$ & Methods & Corrected or not & quasi-normal modes $\omega/M$ & corrections\\
\hline
\multirow{4}{*}{$l_1=1$} & WKB & NO & 0.350922-0.0971917\I
 &\multirow{2}{*}{Almost 0}\\ \cline{2-4}
 & WKB & YES & 0.350922-0.0971915\I
 &\\ \cline{2-5}
 & Prony & NO  & 0.352224-0.0967875\I &\multirow{2}{*}{Almost 0}\\ \cline{2-4}
 & Prony & YES & 0.352218-0.0967840\I &  \\ \hline
 \multirow{4}{*}{$l_2=10$} & WKB & NO & 2.43173-0.0963221\I
 &\multirow{2}{*}{Almost 0}\\ \cline{2-4}
 & WKB & YES & 2.43173-0.0963221\I
 &  \\ \cline{2-5}
 & Prony & NO & 2.43211-0.0962772\I &\multirow{2}{*}{Almost 0}\\ \cline{2-4}
 & Prony & YES &  2.43211-0.0962772\I & \\ \hline
    \end{tabular}
    \caption{Result 5: $M=10~\mathrm{kg},q=0.9$}
\end{table}

\begin{table}[htb!]
    \centering
    \begin{tabular}{|c|c|c|c|c|}
        \hline
Quantum number $l$ & Methods & Corrected or not & quasi-normal modes $\omega/M$ & corrections\\
\hline
\multirow{4}{*}{$l_1=1$} & WKB & NO & 0.375675-0.0893758\I
 &\multirow{2}{*}{0.067656+2.904$\times 10^{-4}$\I}\\ \cline{2-4}
 & WKB & YES & 0.443331-0.0890854\I
 & \\ \cline{2-5}
 & Prony & NO  & 0.377334-0.0894164\I & \multirow{2}{*}{0.067288-1.560$\times 10^{-4}$\I}\\ \cline{2-4}
 & Prony & YES & 0.444622-0.0892604\I &  \\ \hline
 \multirow{4}{*}{$l_2=10$} & WKB & NO & 2.62510-0.0884262\I
 &\multirow{2}{*}{0.52456+1.32$\times 10^{-5}$\I}\\ \cline{2-4}
 & WKB & YES & 3.14966-0.0884130\I
 &  \\ \cline{2-5}
 & Prony & NO & 2.62558-0.0883785\I & \multirow{2}{*}{0.52490+3.45$\times 10^{-5}$\I}\\ \cline{2-4}
 & Prony & YES & 3.15048-0.0883440\I  &  \\ \hline
    \end{tabular}
    \caption{Result 6: $M=10~\mathrm{kg},q=0.9999$}
\end{table}

\begin{table}[htb!]
    \centering
    \begin{tabular}{|c|c|c|c|c|}
        \hline
Quantum number $l$ & Methods & Corrected or not & quasi-normal modes $\omega/M$ & corrections\\
\hline
\multirow{4}{*}{$l_1=1$} & WKB & NO & 0.350922-0.0971917\I &\multirow{2}{*}{1.6$\times10^{-5}$+3$\times10^{-7}$\I}\\ \cline{2-4}
 & WKB & YES & 0.350938-0.0971914\I &  \\ \cline{2-5}
 & Prony & NO  & 0.352224-0.0967875\I & \multirow{2}{*}{1.7$\times 10^{-5}$-2.86$\times 10^{-5}$\I}\\ \cline{2-4}
 & Prony & YES & 0.352241-0.0968161\I  & \\ \hline
 \multirow{4}{*}{$l_2=10$} & WKB & NO & 2.43173-0.0963221\I
 &\multirow{2}{*}{1.3$\times 10^{-4}$+7$\times 10^{-7}$\I}\\ \cline{2-4}
 & WKB & YES & 2.43186-0.0963214\I
 & \\ \cline{2-5}
 & Prony & NO & 2.43211-0.0962772\I & \multirow{2}{*}{1.3$\times 10^{-4}$+7$\times 10^{-7}$\I} \\ \cline{2-4}
 & Prony & YES & 2.43224-0.0962765\I &  \\ \hline
    \end{tabular}
    \caption{Result 7: $M=1~\mathrm{kg},q=0.9$}
\end{table}

\begin{table}[htb!]
    \centering
    \begin{tabular}{|c|c|c|c|c|}
        \hline
Quantum number $l$ & Methods & Corrected or not & quasi-normal modes $\omega/M$ & corrections\\
\hline
\multirow{4}{*}{$l_1=1$} & WKB & NO & 0.375675-0.0893758\I
 &\multirow{2}{*}{2.001312+9.681$\times 10^{-4}$\I}\\ \cline{2-4}
 & WKB & YES & 2.376987-0.0884077\I &  \\ \cline{2-5}
 & Prony & NO  & 0.377334-0.0894164\I & \multirow{2}{*}{2.000015+1.0478$\times 10^{-3}$\I}\\ \cline{2-4}
 & Prony & YES & 2.377349-0.0883687\I & \\ \hline
 \multirow{4}{*}{$l_2=10$} & WKB & NO & 2.62510-0.0884262\I
 &\multirow{2}{*}{14.97624+4.32$\times 10^{-5}$\I}\\ \cline{2-4}
 & WKB & YES & 17.60134-0.0883834\I &  \\ \cline{2-5}
 & Prony & NO & 2.62558-0.0883784\I & \multirow{2}{*}{15.12097+2.2727$\times 10^{-3}$\I}\\ \cline{2-4}
 & Prony & YES & 17.74655-0.0861057\I &  \\ \hline
    \end{tabular}
    \caption{Result 8: $M=1~\mathrm{kg},q=0.9999$}
\end{table}
From our results, the WKB and Prony methods yield very close results, indicating that the quantum-gravity-corrected quasi-normal modes differ from the classical ones.
\end{widetext}

\clearpage
\subsection{Discussion of the Results}

In our calculation, we found that the quantum gravity correction affects the real part more strongly than the imaginary part. Here, we take the 1st-order WKB approximation to show why this is true.

Recall that from Eq.~\eqref{eq:WKB}, we can write the quasi-normal modes of as
\begin{equation}
    \omega=\sqrt{V_0-i\left(n+\frac{1}{2}\right)\sqrt{-2V''(r_0)}}\,,
\end{equation}
where $V_0$ denotes the maximum of the effective potential. From our results, it can be seen that the imaginary part is much smaller than the real part, so we can use the Taylor expansion to approximate the quasi-normal modes as:
\begin{equation}
    \omega=\sqrt{V_0}-\frac{i}{2}\left(n+\frac{1}{2}\right)\sqrt{\frac{-2V''(r_0)}{V(r_0)}}\,.
\end{equation}
The quantum-corrected effective potential is given in Eq.~\eqref{eq: potential} as
\begin{equation}
    V(r)=f(r)\left(\frac{l(l+1)}{r^2}\langle A_{\mathrm{cor}}\rangle+\frac{f'(r)}{r}\right)\,,
\end{equation}
from which we can obtain the quantum-corrected quasi-normal modes.

Numerical calculations show that the quantum correction shifts the potential's maximum by a very slight amount. In the range around the potential peak, the quantum correction is equivalent to multiplying the original potential by a scale factor.
We can understand it as follows. Since we know that $f(r)=1-2M/r+Q^2/r^2$, the potential can be written as
\begin{widetext}
    \begin{equation}
        V(r)=\left(1-\frac{2M}{r}+\frac{Q^2}{r^2}\right)\left(\frac{l(l+1)}{r^2}\langle A_{\mathrm{cor}}\rangle+\frac{2M}{r^3}-\frac{2Q^2}{r^4}\right)=\frac{l(l+1)}{r^2}\langle A_{\mathrm{cor}}\rangle+O(r^{-3})\,,
\end{equation}
\end{widetext}
where we have performed an expansion for large $r$. This expansion is legitimate because the potential maximum is always outside the horizon, i.e., $r>M=1$ in our units.

Meanwhile, we can numerically calculate the potential's maximum value and its second derivative at the maximum. The results are shown in the following table.
\begin{widetext}
    \begin{table}[htb!]
    \centering
    \begin{tabular}{|c|c|c|c|c|c|c|c|c|}
        \hline
        Mass & 1 & 1 & 10 & 10 & 100 & 100 & 1000 & 1000\\
        \hline
        Charge & 0.9 & 0.9999 & 0.9 & 0.9999 & 0.9 & 0.9999 & 0.9 & 0.9999 \\
        \hline
        Maximum of the potential (without correction) & 0.13757 & 0.15623 & 0.13757 & 0.15623 & 0.13757 & 0.15623 & 0.13757 & 0.15623 \\
        \hline
        Maximum of the potential (with correction) & 0.13758 &5.6638 & 0.13757 &0.21130 & 0.13757 & 0.15678 & 0.13757 & 0.15623 \\
        \hline
        Second derivative at maximum (without correction) & -0.14874 & -0.17188 &-0.14874 &-0.17188 & -0.14874 & -0.17188 & -0.14874 & -0.17188 \\
        \hline
        Second derivative at maximum (with correction) & -0.14875 & -5.6762 &-0.14874 &-0.22691 & -0.14874 & -0.17241 & -0.14874 & -0.17186 \\
        \hline
    \end{tabular}
    \caption{Potential's maxima and second derivatives at maxima $(l=1)$}
    \label{tab: potential at maxima}
\end{table}
\end{widetext}

 From these results, we can see that near the potential's maximum, the quantum correction acts as a scale multiplication, under which the real part, $\sqrt{V_0}$, does get a significant correction. In contrast, the imaginary part, proportional to $\sqrt{-V''/V}$, is almost invariant under quantum correction.

\clearpage

\newpage

\bibliography{ref}

\end{document}